\documentclass[11pt]{article}
\usepackage{hyperref}
\usepackage{authblk}
\usepackage{graphicx}
\usepackage{caption}
\usepackage{amsmath}
\usepackage{amssymb}
\usepackage{mathtools}
\usepackage{multirow}
\usepackage{amsmath}
\usepackage{color}
\usepackage{subcaption}
\usepackage{wasysym}
\usepackage{ulem}

\begin{document}

\title{Low energy nuclear reactions in a crystal lattice}

\author{Harishyam Kumar$^{1}$\footnote{hari@iitk.ac.in}}
\author{Pankaj Jain$^{2}$\footnote{pkjain@iitk.ac.in}}
\author{K. Ramkumar$^{1}$\footnote{ramkumar@iitk.ac.in}}
\affil{
$^1$ Physics Department, Indian Institute of Technology, Kanpur, 208016, India\\
$^2$ Department of Space Science \& Astronomy, Indian Institute of Technology, Kanpur, 208016, India}

\maketitle

\begin{abstract}
We extend the recently proposed mechanism for inducing low energy nuclear reactions (LENRs) to a crystal lattice. The process gets dominant contribution at
second order in quantum perturbation theory. The tunnelling barrier is 
evaded due to the presence of high energy intermediate states. However,
the process is found to have negligible rate in free space due a delicate
cancellation between amplitudes corresponding to different states. The delicate
cancellation can be evaded in medium in special circumstances leading
to an observable rate. Here we consider this process in a one 
	dimensional crystal lattice
and find that the rate depends strongly on the form of interaction. 
We find that for some allowed choice of interactions the rate is substantial
	while other interaction produce very small rate. 
\end{abstract}

\section{Introduction}

There exists considerable experimental evidence for nuclear fusion reactions
at low energies \cite{doi:10.1002/9781118043493.ch41,doi:10.1002/9781118043493.ch42,doi:10.1002/9781118043493.ch43,StormsCS2015,mckubre2015cold,articleBi,SRINIVASAN2020233,Mizuno19}.
However, theoretically it is so far not clear how such reactions may occur, 
although there exist many proposals \cite{krivit2009new,SinhaCS2015,LiangCS2015,Hagelstein15,Celani17,kalman2019low,SPITALERI2016275,Meulenberg19,PhysRevC.99.054620} 
In a recent series of papers \cite{Jain2020, Jain2021, Ramkumar2022} 
 we have suggested that low energy nuclear reactions may get significant 
 contribution at second order in perturbation theory. 
 This is 
 similar to the mechanism 
proposed in \cite{PhysRevC.99.054620}. 
 This is illustrated
 by the basic formula at this order \cite{merzbacher1998quantum,sakurai1967advanced}
\begin{equation}
    \langle f|T(t_0,t)|i\rangle = \left( -{i\over \hbar}\right)^2\sum_m \int_{t_0}^t dt' e^{i(E_f-E_m)t'/\hbar} \langle f|H_I(t')|m\rangle \int_{t_0}^{t'} dt'' e^{i(E_m-E_i)t''/\hbar} \langle m|H_I(t'')|i\rangle
    \label{eq:Transition} \ 
\end{equation}
where $|i\rangle$ is an initial low energy state, $|f\rangle$ is the final
state, $H_I(t)$ is the interaction Hamiltonian and $|m\rangle$ represent the
intermediate states. Here $E_i$, $E_m$ and $E_f$ represent the initial, 
intermediate and final state energies and we need to sum over all the 
intermediate state energy eigenvalues. 
The process being considered is different from those generally considered
at first order which are expected to be strongly suppressed at low
energies \cite{1968psen.book.....C}. At second order new processes open 
up which do not get any contribution at first order. In \cite{Jain2020,Jain2021}, for example, the authors considered process with emission of two photons
which necessarily requires two perturbations and hence vanishes at first order. 

The process takes place in two steps. 
At the first interaction vertex the
initial state $|i\rangle$ makes a transition to an intermediate state $|m\rangle$. At the second interaction vertex the 
intermediate state $|m\rangle$ forms the final state nuclei. At both
the vertices a additional particle may be emitted or some energy/momentum
transfered
to the lattice or an impurity particle in the medium. 
The important point is that
the intermediate energies can take arbitrarily large values and energy need 
not be conserved at each interaction vertex. Hence the Coulomb barrier,  
may not lead to a significant suppression and the amplitude $\langle f|H_I(t')|m\rangle $ which leads to the fusion process may not be too small. 

In our earlier calculations \cite{Jain2020, Jain2021} we assumed that the interaction vertices are 
electromagnetic leading to emission of a photon at each vertex. A direct
calculation reveals that, although for some intermediate state energies $E_m$
the amplitude becomes rather large, the sum over all $E_m$ takes very small
values. Hence the rate turns out to be very small. The basic problem arises
at the first vertex. Here we are dealing with eigenstates in a potential
which are not momentum eigenstates. We find that a large number of these
states contribute and mutually cancel the contribution due to one another. 
This problem would not have arisen if we had momentum eigenstates and 
imposed momentum conservation at this vertex. However this is not applicable 
in the model used in \cite{Jain2020,Jain2021}. 

In \cite{Ramkumar2022}, we studied this problem in more detail by using a
toy model for the potential. We used a step model for the potential barrier
since it admits simple analytic expressions for the wave functions. We found
that if we assume free space wave functions at large distances, the reaction
rate is very small due to a delicate cancellation among different wave 
functions. This is in agreement with results found in \cite{Jain2020,Jain2021}.
However, in a medium the energy eigenvalues are expected to
be discretized. This is true both for a crystalline lattice \cite{Ashcroft76}
and for disordered systems \cite{doi:10.1142/7663,RevModPhys.57.287}. 
This idea was implemented in a simple manner in \cite{Ramkumar2022}
by imposing an infinite potential wall at a large distance from the nuclear
potential well. This leads to discretization of energy eigenvalues and now
the cancellation is not exact. In this case we find that the reaction rate
can be substantial, much larger than rate for the first order process which is
strongly suppressed by the potential barrier due to low initial state 
energy. This shows that although the rate is negligible in free space it 
can be substantial in a medium. This is also consistent with experimental 
claims \cite{doi:10.1002/9781118043493.ch41,doi:10.1002/9781118043493.ch43,StormsCS2015,Biberian19}, which suggest that the process happens only in a medium. 

\begin{figure}
     \centering
     \includegraphics[width=0.84\textwidth]{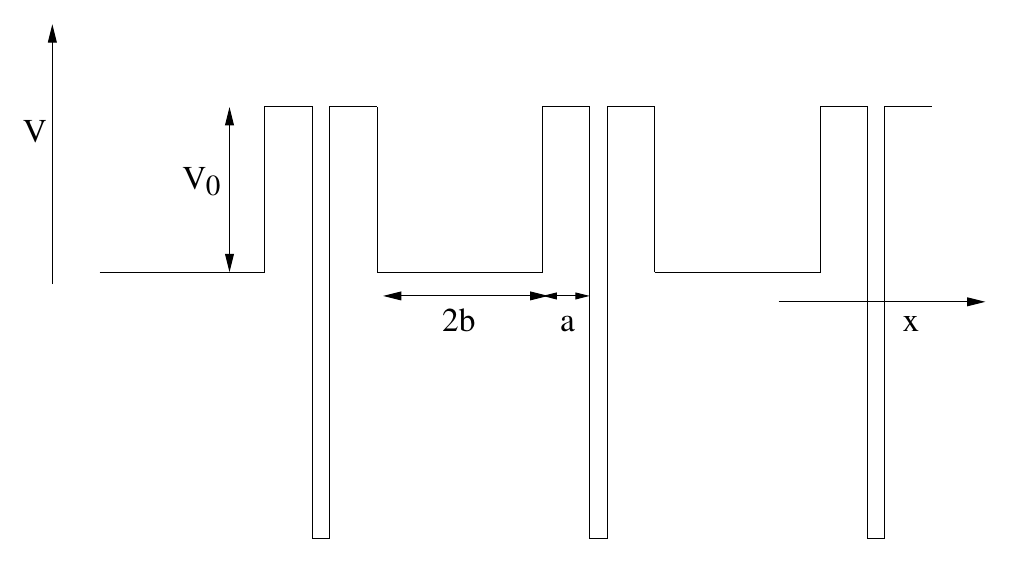}\\
     \caption{Schematic illustration of the periodic potential given
	in Eq. \ref{eq:perpotential}. 
}
\label{fig:potential}
\end{figure}

In this paper we consider the fusion process in a crystal lattice, 
which provides a realistic model for solid medium.
We set up the formalism in general but restrict our calculations to 
the one dimensional Kronig-Penney model \cite{merzbacher1998quantum}.
In this case a proton is assumed to undergo fusion with a lattice ion. 
We assume
that the proton experiences 
a periodic potential such that
\begin{equation}
	V(\vec r + \vec R) = V(\vec r)
	\label{eq:potential}
\end{equation}
We emphasize that we 
are interested in the potential experienced by a proton and not an
electron. Hence at the location of each ion we expect a strong repulsive potential and an attractive potential between the sites. 
The form of the potential is similar to that of electrons
but we expect a much stronger
barrier height in the present case. 
A sample potential in one dimension, which is taken to be a generalization of 
the Kronig-Penney potential \cite{merzbacher1998quantum}, is
illustrated in Fig. \ref{fig:potential}.
The deep and narrow attractive potential models the 
nuclear force. 
This leads to a modification of the wave function at distances very close to
the nucleus.
At large distances it is expected to lead to only a mild
change. For example, in free space it leads to phase shift which shows a 
mild dependence on energy. 
The potential 
 can be expressed as, 
 \begin{eqnarray}
        V(x) &=& 0\ \ \ \ \ {\rm for} \ \ a+(n-1)\xi <x<n\xi - a\nonumber\\
             &=& V_0 \ \ \ \ {\rm for} \ \ n\xi -a < x< n\xi - d\nonumber\\
             &=& -V_1 \ \ \ \ {\rm for} \ \ n\xi -d < x< n\xi + d\nonumber\\
             &=& V_0 \ \ \ \ {\rm for} \ \ n\xi +d < x< n\xi + a
             \label{eq:perpotential}
\end{eqnarray}
where $\xi = 2a+2b$.
Here $d$ is very small and represents
the nuclear length scale. Furthermore the nuclear
potential $V_1$ is taken to be very strong.

The eigenfunctions in a periodic lattice can be expressed as
\begin{equation}
	\psi(\vec r) = {1\over \sqrt {N{\cal N}}} 
	e^{i\vec k\cdot \vec r} u_k(\vec r) 
	\label{eq:protonwf}
\end{equation}
such that 
\begin{equation}
	u_k(\vec r+\vec R) = u_k(\vec r) 
	\label{eq:periodicu}
\end{equation}
Here $N$ is the number of unit cells and ${\cal N}$ is the normalization factor 
in each unit cell. 
We shall consider the process in which the proton 
emits a particle, such as a photon, at the first vertex. The emitted
particle may be a virtual photon which may be absorbed by another particle
in the medium. In earlier papers we have considered emission of real
photons \cite{Jain2020,Jain2021,Ramkumar2022}. In the present paper we perform
an explicit calculation using a one dimensional model. Hence, for simplicity,
we consider emission of a scalar (spin 0) particle since in one dimension
only one spatial component is allowed.

\section{General Formalism}
In this section we develop the general formalism for the second order process.
We shall perform an explicit calculation only in one dimension, however, 
here we develop the framework assuming three dimensions. 
The process involves two perturbations or vertices. At the first vertex, the
incident particle, which we refer to as a 
proton, emits a spin 0 particle of high momentum. 
At the second vertex it undergoes
fusion with a lattice site while emitting another spin 0 particle. As explained
earlier, the emitted particles are taken to be spin 0 since we perform
explicit calculations only in 1 dimension. The net effect is fusion of
proton with a lattice site with emission of two scalar particles.

The interaction Hamiltonian can
be expressed as
\begin{equation}
	H_I = -g\phi(\vec r,t){\cal O} 
\label{eq:HI}
\end{equation}
where $\phi$ is a scalar field and can be expressed as,
\begin{equation}
	\phi(\vec r,t) = {1\over \sqrt{\Omega}} \sum_{\vec q} \sqrt{\hbar\over 2\omega(q)} \left[a_{\vec q}(t) e^{i\vec q\cdot\vec r} +  a^\dagger_{\vec q}(t) 
 e^{-i\vec q\cdot\vec r}\right] \,. 
	\label{eq:Hint}
\end{equation}
 Here $\omega(q)$ is the frequency of spin 0 particles, $\Omega$ the spatial volume
and $a(\vec q)$ the annihilation operator. 
In Eq. \ref{eq:HI}, ${\cal O}$ is an operator which can be chosen to be an 
identity or some other operator. We discuss its detailed form later. 
The form of $\phi$ is chosen
so that it has the right dimensions in one spatial dimension since 
our detailed calculation will be done only for this case.
The interaction Hamiltonian in Eq. \ref{eq:HI} can have dependence on the
momentum of the incident proton in analogy
to the interaction with electromagnetic field \cite{Jain2020,Jain2021}. 
This is implemented by introduction of the operator ${\cal O}$.
 The constant $g$ has dimensions
which will be specified later.

We next 
compute the matrix element $\langle m|H_I(t^{\prime\prime})|i\rangle$ in Eq. \ref{eq:Transition} using the interaction Hamiltonian in Eq. \ref{eq:HI} setting
the operator ${\cal O}$ equal to unity. 
This perturbation leads to production of a 
 scalar particle of momentum $\vec q_1$ and energy $E_1=\hbar \omega_1$. 
We obtain,
\begin{equation}
	\langle m|H_I(t^{\prime\prime})|i\rangle = -{g\over \sqrt{\Omega}}
	\sqrt{\hbar\over 2\omega_1} \int d^3 r \psi_m^*(\vec r) 
	e^{-i\vec q_1\cdot \vec r}e^{i(E_m+\hbar\omega_1-E_i) t^{\prime\prime}/\hbar}
	 \psi_i(\vec r)
\end{equation}
where $\psi_i$ and $\psi_m$ represent the initial and intermediate state
wave function corresponding to wave vectors $\vec k_i$ and $\vec k_m$ 
respectively.
This leads to 
\begin{equation}
\langle m|H_I(t^{\prime\prime})|i\rangle = -{g\over \sqrt{\Omega {\cal N}_i{\cal N}_m}N}
	\sqrt{\hbar\over 2\omega_1}
	e^{i(E_m+\hbar\omega_1-E_i) t^{\prime\prime}/\hbar}	
	\int d^3 r e^{-i\vec k_m\cdot \vec r} u^*_{k_m} (\vec r) 
	e^{-i\vec q_1\cdot \vec r}  
	 e^{i\vec k_i\cdot \vec r} u_{k_i} (\vec r)
\end{equation}
Here we have also substituted for the proton wave functions using Eq. 
\ref{eq:protonwf} and ${\cal N}_i$ and ${\cal N}_m$ are the normalization
factors for the initial and intermediate states respectively. 
We now split the integral into an integral over a unit cell and sum 
over all cells. Hence we write $\vec r$ as
\begin{equation}
\vec r = \vec s + \vec R_j
\end{equation}
This leads to,
\begin{eqnarray}
	\langle m|H_I(t^{\prime\prime})|i\rangle &=& -{g\over \sqrt{\Omega {\cal N}_i{\cal N}_m}N}
	\sqrt{\hbar\over 2\omega_1}
	e^{i(E_m+\hbar\omega_1-E_i) t^{\prime\prime}/\hbar}\nonumber\\	
	&\times &\sum_j\int d^3 s e^{-i\vec k_m\cdot (\vec s+\vec R_j)} u^*_{k_m} (\vec s) 
	e^{-i\vec q_1\cdot (\vec s+\vec R_j)}  
	 e^{i\vec k_i\cdot (\vec s+\vec R_j)} u_{k_i} (\vec s)
\end{eqnarray}
where we have used the periodicity of $u_k$ displayed in Eq. \ref{eq:periodicu}.
We can now perform the sum over $j$ using,
\begin{equation}
	\sum_j e^{i(-\vec k_m - \vec q +\vec k_i)\cdot R_j} = N \sum_{\vec G}
	\delta_{\vec G,-\vec k_m - \vec q_1 +\vec k_i}
\end{equation}
where $\vec G$ are the reciprocal lattice vectors. Substituting this, we obtain,
\begin{eqnarray}
	\langle m|H_I(t^{\prime\prime})|i\rangle &=& -{g\over 
	\sqrt{\Omega {\cal N}_i{\cal N}_m}}
	\sqrt{\hbar\over 2\omega_1}
	e^{i(E_m+\hbar\omega_1-E_i) t^{\prime\prime}/\hbar}\nonumber\\	
&\times & \sum_{\vec G} \delta_{\vec G,-\vec k_m - \vec q +\vec k_i}\int d^3 s 
	e^{-i\vec k_m\cdot \vec s} u^*_{k_m} (\vec s) 
	e^{-i\vec q_1\cdot \vec s}  
	 e^{i\vec k_i\cdot \vec s} u_{k_i} (\vec s)
\end{eqnarray}
Hence $\vec k_m$ takes only those values which satisfy,
 $\vec k_m = -\vec q_1+\vec k_i - \vec G$. 

We next consider the second matrix element
$\langle f|H_I(t^\prime)|m\rangle$. In this case the intermediate state
makes a transition to final state with emission of another scalar
particle of wave vector $\vec q_2$ and energy $E_2= \hbar \omega_2$. The
proton undergoes fusion with a lattice ion, whose atomic number changes
from $Z$ to $Z'=Z+1$. Hence the final state is the proton bound to a
lattice ion due to the nuclear potential which is the narrow deep well
shown in Fig. 
\ref{fig:potential}. 
This wave function is also 
periodic and takes large values only very close to an ion over the
range of nuclear potential and decays
rapidly between two ions. Hence the final state wave function also takes  
the form shown in Eq. \ref{eq:protonwf} and \ref{eq:periodicu}. 
Physically this implies that the proton can undergo fusion with any of
the lattice ions.

We obtain
	\begin{equation}
\langle f|H_I(t^\prime)|m\rangle  
		= -{g\over \sqrt{\Omega {\cal N}_f{\cal N}_m }N}
        \sqrt{\hbar\over 2\omega_2}
		e^{i(E_f+\hbar\omega_2-E_m) t^{\prime}/\hbar}
        \int d^3 r e^{-i\vec k_f\cdot \vec r} u^*_{k_f} (\vec r)
        e^{-i\vec q_2\cdot \vec r}
         e^{i\vec k_m\cdot \vec r} u_{k_m} (\vec r)
\end{equation}
Here we have replaced the wave functions 
using Eq. \ref{eq:protonwf} and ${\cal N}_f$ is the normalization
factor corresponding to the final state. We point out that the 
wave vector $\vec k_f$ depends
on the value of the deep nuclear potential well. We again 
set $\vec r = \vec s+\vec R_l$ and use periodicity to replace the integral in
terms of integral over a unit cell and sum over all cells. In this case the
sum over cells leads to
\begin{equation}
	\sum_l e^{i(-\vec k_f - \vec q_2 +\vec k_m)\cdot R_l} = N 
	\sum_{\vec G'}
	\delta_{\vec G',-\vec k_f - \vec q_2 +\vec k_m}
\end{equation}
We need to compute the matrix element for a fixed value of
the final state wave vector $\vec k_f$. For the fixed value of this wave vector, the wave vector of the final state $\phi$ particle takes the value 
$\vec q_2= -\vec k_f + \vec k_m -\vec G'=\vec k_i - \vec G'-\vec G -\vec q_1-\vec k_f$ 
where $\vec G'$ is a reciprocal lattice vector. 
Hence we obtain,
	\begin{eqnarray}
		\langle f|H_I(t^\prime)|m\rangle &=& 
		-{g\over \sqrt{\Omega {\cal N}_f{\cal N}_m}}
        \sqrt{\hbar\over 2\omega_2}
	e^{i(E_f+\hbar\omega_2-E_m) t^{\prime}/\hbar}\nonumber\\
&\times &	\sum_{\vec G'}
	\delta_{\vec G',-\vec k_f - \vec q_2 +\vec k_m}
        \int d^3 s e^{-ik_f\cdot \vec s} u^*_{k_f} (\vec s)
        e^{-i\vec q_2\cdot \vec s}
         e^{ik_m\cdot \vec s} u_{k_m} (\vec s)
\end{eqnarray}
In this equation the integral is only over a unit cell. 

The transition amplitude can now be expressed as
\begin{equation}
\langle f|T(t_0,t)|i\rangle = 
	{i\over \hbar} 
	{g^2\hbar \over 2 \Omega \sqrt{\omega_1\omega_2}}
	\int_{t_0}^t dt' e^{i(E_f+\hbar \omega_2+\hbar \omega_1 - E_i)t'/\hbar} I
	\label{eq:MatFI}
\end{equation}
where 
\begin{equation}
	I = 
	\sum_m \sum_{\vec G'} \sum_{\vec G}         
	\delta_{\vec G',-\vec k_f - \vec q_2 +\vec k_m}
  {I_2 I_1\over E_m-E_i+\hbar \omega_1} \delta_{\vec G,-\vec k_m - \vec q_1 +\vec k_i}
\label{eq:IntI}
\end{equation}
\begin{equation}
I_1 =	\int d^3 s  \psi^*_{k_m} (\vec s)
        e^{-i\vec q_1\cdot \vec s}
         \psi_{k_i} (\vec s)
	 \label{eq:I1}
\end{equation}
and
\begin{equation}
I_2 = \int d^3 s \psi^*_{k_f} (\vec s)
        e^{-i\vec q_2\cdot \vec s}
          \psi_{k_m} (\vec s)
	 \label{eq:I2}
\end{equation}
Here we have absorbed the normalization factors in the wave functions. 
We may use either the extended or the reduced band scheme. In the extended
band scheme each $k_m$ is uniquely associated with an energy eigenvalue
$E_m$. 

For some values of $\vec k_i$, $\vec k_m$ and $\vec q_1$, the matrix element
is expected to be quite large. However, in analogy with the 
results in \cite{Jain2020,Jain2021}, it is possible that as we sum the 
complete amplitude over all $k_m$ the amplitude may cancel out. 
As seen in \cite{Ramkumar2022}, such a cancellation does not happen if
the energy spectrum takes discrete values. This may be applicable in the
present case also. 
In the sum over $m$, $\vec k_m$ take only those values which satisfy
$\vec k_m = \vec q_1+\vec k_1-\vec G$ and the energies are corresponding
restricted. This leads to discretization of energy eigenvalues since in each
band we get contribution from only one state.
Hence we may find a substantial rate in the present case also.
We examine this by using
 the generalized one dimensional
Kronig-Penney model in the next section.

\section{One dimensional generalized Kronig-Penney Model}
We specialize to the one-dimensional potential given in Eq. \ref{eq:perpotential}. The eigenfunctions for $E<V_0$ are given by,
\begin{eqnarray}
	\psi &=&{1\over{\cal N}}  \left[ A_n e^{ik'(x-n\xi)} + B_n e^{-ik'(x-n\xi)}\right]  \ \ \ \ \ {\rm for} \ \ a+(n-1)\xi <x<n\xi - a\nonumber\\  
	 &=&{1\over{\cal N}}\left[ C_n e^{-\kappa(x-n\xi)} + D_ne^{\kappa(x-n\xi)}\right] \ \ \ \ {\rm for} \ \ n\xi -a < x< n\xi -d\nonumber\\ 
	 &=&{1\over{\cal N}}\left[ E_n e^{iQ(x-n\xi)} + F_ne^{-iQ(x-n\xi)}\right] \ \ \ \ {\rm for} \ \ n\xi -d < x< n\xi +d\nonumber\\ 
	 &=& {1\over{\cal N}} \left[ G_n e^{-\kappa(x-n\xi)} + H_ne^{\kappa(x-n\xi)}\right] \ \ \ \ {\rm for} \ \ n\xi +d < x< n\xi +a
\end{eqnarray}
where
\begin{equation}
        k' = \sqrt{2mE\over \hbar^2}\ ,
	\label{eq:kprime}
\end{equation}
\begin{equation}
        \kappa = \sqrt{2m(V_0-E)\over \hbar^2}\ ,
\end{equation}
\begin{equation}
        Q = \sqrt{2m(E+V_1)\over \hbar^2}\ ,
\end{equation}
$m$ is the mass of the particle,
$E$ is the energy eigenvalue and ${\cal N}$ is the normalization factor.
It is convenient to work in analogy with the standard Kronig-Penney model 
\cite{merzbacher1998quantum}.
Here we use the same formalism and notation.
The coefficients $A_n$ in two adjacent cells are related by
\begin{equation}
	\begin{pmatrix}A_n\\ B_n   
	\end{pmatrix}  
	= \begin{pmatrix} \alpha_1+i\beta_1 & i\beta_2\\
		-i\beta_2 & \alpha_1-i\beta_1
	\end{pmatrix} \begin{pmatrix} A_{n+1}e^{-ik'\xi}\\
		B_{n+1} e^{ik'\xi} 
	\end{pmatrix}
\end{equation}
The variables $\alpha_1$, $\beta_1$ and $\beta_2$ are given
by
\begin{eqnarray}
	\alpha_1 &=& {\cos 2k'a\over 2}\left[
		V\cosh 2\kappa(a-d) - X\sinh2\kappa(a-d) 
	\right] \nonumber\\
	&-& {\sin 2k'a\over 4}
	\Bigg\lbrace\left({k'\over \kappa} - {\kappa\over k'}\right)
	\left[ -V\sinh 2\kappa(a-d)+  X\cosh2\kappa(a-d)\right] - Y
	\left( {k'\over \kappa} + {\kappa\over k'} \right) \Bigg\rbrace\\
	\beta_1 &=& {\sin 2k'a\over 2}\left[
	V\cosh 2\kappa(a-d) -
		X\sinh 2\kappa(a-d)  
	\right] \nonumber\\
	&+& {\cos 2k'a\over 4}
	\Bigg\lbrace\left({k'\over \kappa} - {\kappa\over k'}\right)
	\left[ -V\sinh2\kappa(a-d)+  X\cosh2\kappa(a-d)	\right] - Y
	\left( {k'\over \kappa} + {\kappa\over k'} \right) \Bigg\rbrace \\
	\beta_2 &=& {1\over 4}\Bigg\lbrace \left( {k'\over \kappa} + {\kappa\over k'} \right) \left[V\sinh 2\kappa(a-d) - X\cosh2\kappa(a-d)\right]
	+ Y \left({k'\over \kappa} - {\kappa\over k'}\right)\Bigg\rbrace 
\end{eqnarray}
where
\begin{eqnarray}
	V &=&2\cos 2Qd\nonumber\\
	X &=& \left({Q\over \kappa} - {\kappa\over Q}\right)\sin2Qd\nonumber\\
	Y &=& \left({Q\over \kappa} + {\kappa\over Q}\right)\sin2Qd \\
\end{eqnarray}
The corresponding quantities for $E>V_0$ are obtained by replacing $\kappa$
with $-iK$ where
\begin{equation}
        K = \sqrt{2m(E-V_0)\over \hbar^2}\ ,
\end{equation}

We define the transfer operator $P$ \cite{merzbacher1998quantum}, 
\begin{equation}
P = \begin{pmatrix} (\alpha_1-i\beta_1)e^{ik'\xi} & -i\beta_2 e^{ik'\xi}\\
	i\beta_2e^{-ik'\xi} & (\alpha_1+i\beta_1)e^{-ik'\xi}
\end{pmatrix}
\end{equation}
which satisfy $\det P=\alpha_1^2 + \beta_1^2 - \beta_2^2 = 1$. 
The eigenfunctions take the form of Eq. \ref{eq:protonwf} 
with 
\begin{equation}
\cos k\xi = \alpha_1 \cos k'\xi + \beta_1 \sin k'\xi
	\label{eq:eigenvalue_cond}
\end{equation}
For any choice of $k$ this leads to all the allowed energy eigenvalues. 
The eigenvectors are obtained by solving the matrix equation
\begin{equation}
P\begin{pmatrix}A_0^\pm\\ B_0^\pm   
\end{pmatrix} = p_\pm \begin{pmatrix}A_0^\pm\\ B_0^\pm \end{pmatrix}
\end{equation}
	where the eigenvalues $p_\pm = \exp(\pm ik\xi)$. 

We next compute the integrals $I_1$ (Eq. \ref{eq:I1}) 
and $I_2$ (Eq. \ref{eq:I2}) 
using the generalized
Kronig-Penney model for eigenvalue $p_+$ of the transfer operator. 
Here we specialize to one cell and set $n=0$. 
The coefficients $A_0^+$ and $B_0^+$ can be written as,
\begin{eqnarray}
	A^+_0 &=& \beta_2 e^{i k'\xi/2}\nonumber\\
	B^+_0 & = & B^+_{0a} e^{-i k'\xi/2}
\end{eqnarray}
where, 
	\begin{equation}
		B^+_{0a} = \alpha_1\sin k'\xi -\beta_1\cos k'\xi - \sin k\xi 
	\end{equation}
We also need to make a choice for the operator ${\cal O}$ in Eq. \ref{eq:HI}. So far we have chosen this to be just the identity operator. For this choice,
	$I_1$ can be
written as
\begin{equation}
I_1 =	\int d x  \psi^*_{k_m} ( x)
        e^{-i q_1 x}
         \psi_{k_i} ( x)
\end{equation}
	Let 
$k'_i$ and $k'_m$ denote the wave vectors defined in 
	Eq. \ref{eq:kprime} corresponding to the initial and intermediate
	states respectively. 
	Furthermore we denote variables, such as, $\beta_2$, $B^+_0$ corresponding to the two wave functions as $\beta_{2m}$, $\beta_{2i}$, $B^+_{0m}$, 
	$B^+_{0i}$ etc.
	The dominant contribution to the integral $I_1$ is obtained from the region $-a-2b<x<-a$ where $V(x)=0$.  
The integral in this region, both for $E<V_0$ and $E>V_0$, is found to be equal to 
\begin{eqnarray}
	I_1 &=& {1\over {\cal N}_i{\cal N}_m} 2e^{i(a+b)q_1} \Bigg[ 
	\beta_{2m}\beta_{2i} {\sin b(k'_m-k'_i+q_1)\over k'_m-k'_i+q_1}
	+ \beta_{2m}B^+_{0ia} {\sin b(k'_m+k'_i+q_1)\over k'_m+k'_i+q_1}
	\nonumber\\
	&+&
	 B^+_{0ma}\beta_{2i} {\sin b(k'_m+k'_i-q_1)\over k'_m+k'_i-q_1}
	+ B^+_{0ma}B^+_{0ia} {\sin b(k'_m-k'_i-q_1)\over k'_m-k'_i-q_1}
	\Bigg]
\end{eqnarray}
We notice that this integral can be substantial in kinematic
regions where the argument of any of the sine function, such as $(k'_m-k'_i+q_1)$, vanishes. In other regions the integral is relatively small 
because the wave function $\psi_i$ is strongly suppressed 
due to the potential barrier. Furthermore the enhancement seen in the region
$-a-2b<x<-a$ due to vanishing of the argument of any of the sine functions, 
  does not happen in the remaining regions and hence contributions
from those regions is further suppressed.
In our calculations we explicitly compute the contribution to $I_1$ in the
regions adjacent to dominant region and found it to be negligible. 

The integral $I_2$, which is analogous to the nuclear matrix element, 
	can be written as
\begin{equation}
I_2 =	\int d x  \psi^*_{k_f} ( x)
        e^{-i q_2 x}
         \psi_{k_m} ( x)
\end{equation}
This integral gets dominant contribution from the region $-d<x<d$ and the 
regions adjacent to this. In the adjacent regions, we get contributions
only from values of x very close to $|x|=d$. 
As $x$ moves further away from this region, the
wave function $\psi_{k_f}$ is very
strongly suppressed. 
	In the region $-d<x<d$ we can write the wave function
	$\psi_{k_f}$ as
\begin{equation}
	\psi_{k_f} = {1\over {\cal N}_f} \left[E_{0f} e^{iQ_fx} + F_{0f}e^{-iQ_fx}\right]
\end{equation}
where 
\begin{equation}
        Q_f = \sqrt{2m(E_f+V_1)\over \hbar^2}\ ,
\end{equation}
In the adjacent regions, it is reasonable to approximate the wave function
to be an exponentially falling function which is practically zero once we
substantially far from the region $-d<x<d$. The contribution to the
integral $I_2$ from the dominant region can be written as,
\begin{eqnarray}
	I_2 &=& 2E^*_{0f} E_{0m} {\sin d(Q_f+q_2-Q_m)\over Q_f+q_2-Q_m}
	+ 2 E^*_{0f} F_{0m} {\sin d(Q_f+q_2+Q_m)\over Q_f+q_2+Q_m} \nonumber\\
	 &+& 2F^*_{0f} E_{0m} {\sin d(Q_f-q_2+Q_m)\over Q_f-q_2+Q_m}
	 + 2F^*_{0f} F_{0m} {\sin d(Q_f-q_2-Q_m)\over Q_f-q_2-Q_m}
\end{eqnarray}
where $Q_m= \sqrt{2m(E_m+V_1)/\hbar^2}$.
In our calculations we have included the contribution from adjacent
regions which also give a significant contribution. 

\section{Results}

We choose the following parameters for the potential:
\begin{eqnarray}
	V_0 &=& 100\ {\rm atomic\ units} \nonumber\\
	V_1 &=& 50.0\times 10^6\ {\rm eV} \nonumber \\
	a & = & 0.1\ {\rm atomic\ units}\nonumber \\
	b & = & 1.0\ {\rm atomic\ units}\nonumber \\
	d & = & 5.0\times 10^{-5} \ {\rm atomic\ units}  
\end{eqnarray}
The initial state is chosen to have a very small energy eigenvalue.
For the chosen parameters we obtain an eigen state with energy equal to
$2.68\times 10^{-3}$ in atomic units, which we choose for our calculations.
This energy
eigenvalue $E_i<<V_0$ and hence it shows very little
dependence on $k_i$ within the band. The same is also true for the
final state wave function whose energy eigenvalue $E_f$ lies deep in the 
nuclear potential well. 
We take the final state eigenfunction corresponding to the energy 
eigenvalue $E_f= -10.96$ MeV. This is the second excited state with the 
ground state energy equal to $-45.3$ MeV. 
Given the initial wave number $k_i$ we need to 
sum over all energy eigenvalues $E_m$ for which $k_m= -q_1+k_i-G$ where 
$G$ are the reciprocal lattice vectors. Here $k_i$, and hence $q_1$
can be chosen arbitrarily. Once this choice is made, we need to sum over
all $k_m$ values arising due to different reciprocal lattice vectors.

The reaction rate in 1 dimension can be written as
\begin{equation}
{dP\over dt} = {1\over \Delta T} \int dE_1 dE_2\rho_1\rho_2 \sum_{k_f} |\langle
	f|T(t_0,t)|i\rangle|^2
\end{equation}
where $E_1=\hbar \omega_1$ and $E_2 = \hbar\omega_2$ are the two emitted
particle energies and the factor $\rho_1=\rho_2 = \Omega/(2\pi\hbar c)$. 
The expression for the matrix element is given in Eq. \ref{eq:MatFI}. 
We obtain
\begin{equation}
{dP\over dt} = {g^4\hbar\over 8\pi c^2}\int {dE_2\over E_1 E_2} \sum_{k_f}
	|I|^2
\end{equation}
along with the energy conservation constraint, $E_f+E_2+E_1-E_i=0$. 
We next express $g$ as
\begin{equation}
g = g' {m_ec^{5/2}\over \sqrt{\hbar}}
	\label{eq:dimlesscoupl}
\end{equation}
where $g'$ is a dimensionless coupling and $m_e$ is the electron mass which is equal to unity in atomic units.

We rewrite the equations for the sum/integrals $I$, $I_1$ and $I_2$ in form
suitable for one dimension:
\begin{equation}
	I = 
	\sum_m \sum_{ G'} \sum_{ G}         
	\delta_{ G',- k_f -  q_2 + k_m}
  {I_2 I_1\over E_m-E_i+\hbar \omega_1} \delta_{ G,- k_m -  q_1 + k_i}
\label{eq:IntI1d}
\end{equation}
\begin{equation}
I_1 =	\int d x  \psi^*_{k_m} ( x) e^{-i q_1  x} \psi_{k_i} ( x)
	 \label{eq:I11d}
\end{equation}
and
\begin{equation}
I_2 = \int d x \psi^*_{k_f} ( x) e^{-i q_2  x} \psi_{k_m} ( x)
	 \label{eq:I21d}
\end{equation}
With $k_i$ and $q_1$ fixed $k_m$ takes only one value in a particular band. 
The sum over $G$ amounts to a sum over $k_m$ values across different bands. 
We also have the sum over $k_f$. Here we will restrict ourselves to a single
band and then the sum is over the $k_f$ within the band. 
We point out that the energy $E_f$ is almost independent of $k_f$ within
a band.
With $k_f$ 
restricted to a single band, it is clear that, for fixed $q_2$, the delta function
$\delta_{ G',- k_f -  q_2 + k_m}$ will restrict $k_f$ to a single value. 
The value of $q_2$ needs to consistent with the energy conservation. Since
$E_f$ is almost independent of $k_f$, this can be easily satisfied.
This takes care of all the sums in the rate formula and we are left with
just a single integral over $E_2$. 
As we change the value of $E_2$, the energy $E_1$ will also change
and we will get contribution from different $k_m$ values within a
band. 

Explicit calculations show that although the integral $I_1$ and $I_2$ 
are substantial for some values of $k_m$, after summing over all intermediate
states, the amplitude $I$ takes very small values. This is similar to
what was seen earlier \cite{Jain2020,Jain2021} and also for the case 
of continuous eigenvalues in \cite{Ramkumar2022}. It is somewhat difficult
to get a precise result for this calculation since it is found to be 
within the numerical error in the calculation. Furthermore the integral
undergoes oscillations and to get a final result one need to go to 
very large values of the intermediate energy eigenvalue. This further
inhibits obtaining a precise result and we do not pursue this further.

Our main aim in this paper is to determine if the proposed second
order fusion mechanism is
prohibited or strongly suppressed in Quantum Mechanics. In  
order to address this we test some other operators in the interaction
Hamiltonian. One interesting choice of operator is
\begin{equation}
	{\cal O} = {1\over P^2 - P_0^2}
	\label{eq:operator}
\end{equation}
where $P$ is the momentum operator and $P_0$ is a parameter such
that $E_0= P_0^2/(2m)$ does not belong to the eigen-spectrum of the model. 
We choose $P_0=1$ in atomic units which satisfies the above mentioned 
constraint for the choice of potential parameters. 
This is a well defined Hermitian operator and 
hence allowed by Quantum Mechanics.
It is not clear which physical interaction may lead to this operator.
It may arise if, during the initial stage, the process takes place through
the Hydrogen atom rather than a free proton. In this case the incident 
momentum $P$ is shared both by the proton and the electron through their 
mutual Coulomb interaction. Hence the process would be suppressed by an
additional factor of $1/P^2$, which acts as a form factor for the Hydrogen atom.
This may further enhance the reaction rate since the Hydrogen atom would 
experience a reduced Coulomb barrier.
In any case, our main aim in studying this interaction is to determine the cancellation seen 
earlier is a generic result or happens only for the some interactions. 
Hence, this operator serves our purpose to determine if mathematically Quantum Mechanics
permits enhanced fusion at second order.
We point out that for the first order process, 
the action of this operator will not lead
to any enhancement of rate. 

The action of
the operator given in Eq. \ref{eq:operator}
on a state is facilitated by making an expansion around $P_0$.
For the wave functions under consideration, the action of $P$ is rather 
simple and one can easily sum the resulting series. Hence we obtain, 
setting $\hbar =1$,
\begin{eqnarray}
	{1\over P^2 - P_0^2}e^{\pm ik'x} &=& {1\over k^{\prime 2} - P_0^2}e^{\pm ik'x} \nonumber\\
	{1\over P^2 - P_0^2}e^{\pm\kappa x} &=& {1\over -\kappa^2 - P_0^2}e^{\pm\kappa x} 
\label{eq:operator1}
\end{eqnarray}
Using this we can easily compute all the matrix elements. 
For this model we refer to the integrals corresponding to 
 $I_1$ and $I_2$ as $\bar I_1$ and $\bar I_2$ respectively and
 the sum $I$ (Eq. \ref{eq:IntI1d}) is denoted by $\bar I$.
We obtain
\begin{eqnarray}
\bar I_1 &=&	\int d x  \psi^*_{k_m} ( x) e^{-i q_1  x} {\cal O} \psi_{k_i} ( x)\nonumber\\
\bar I_2 &=& \int d x \psi^*_{k_f} ( x) e^{-i q_2  x} {\cal O} \psi_{k_m} ( x)
	\label{eq:barI1I2}
\end{eqnarray}
and the operation of ${\cal O}$ can be computed using Eq. \ref{eq:operator1}.
Furthermore $\bar I$ is obtained by replacing $I_1$ and $I_2$ in 
Eq. \ref{eq:IntI1d} by $\bar I_1$ and $\bar I_2$ respectively. 
It is also convenient to define a
dimensionless coupling $g''$ for this interaction. In analogy with Eq. \ref{eq:dimlesscoupl}, we obtain,
\begin{equation}
g = g'' {m_e^3c^{9/2}\over \sqrt{\hbar}}
\end{equation}

\begin{figure}
\centering
\includegraphics[ width=0.84\textwidth]{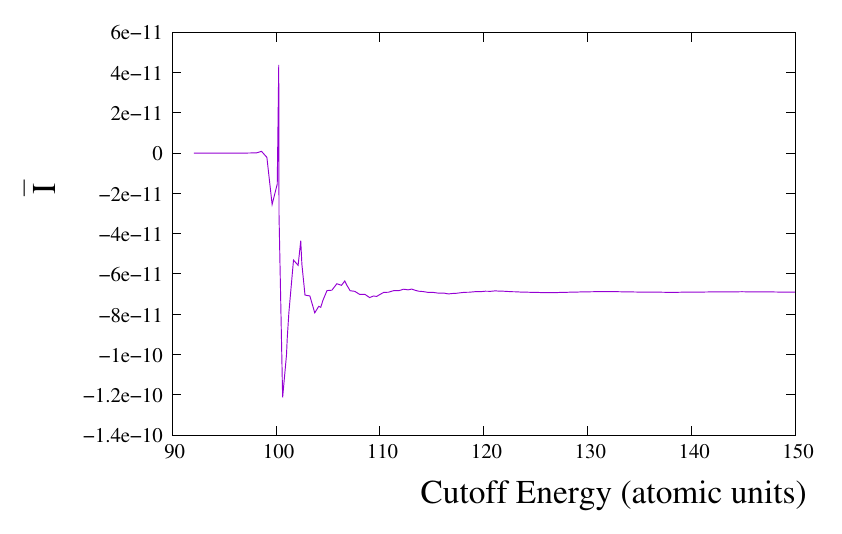}\\ 
	\caption{ The real part of $\bar I$ (see Eq. \ref{eq:barI1I2}) as a 
function of the upper cutoff on the intermediate state energy $E_m$ in 
atomic units. We see that the amplitude tends to a significant
	value as the cutoff goes to infinity.
        }
\label{fig:Ecutoffdep}
\end{figure}

\begin{figure}
\centering
\includegraphics[ width=0.84\textwidth]{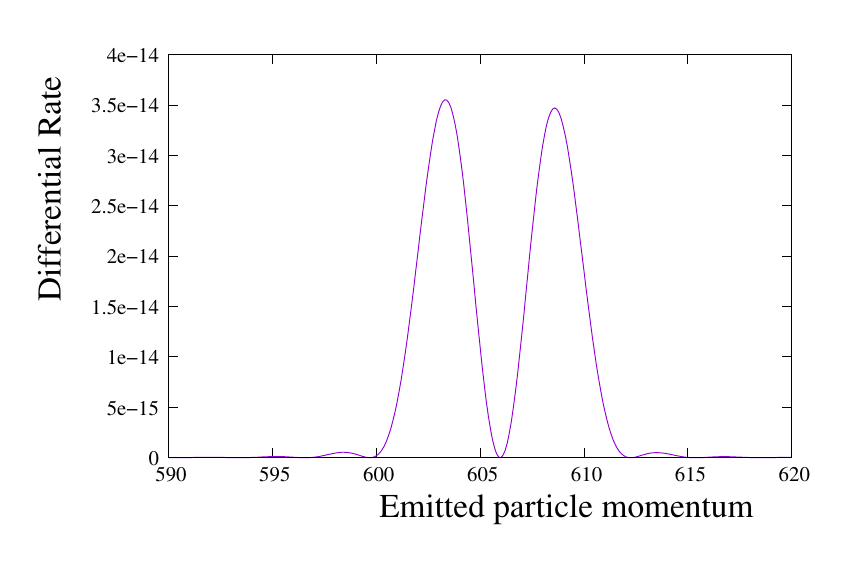}\\ 
\caption{The differential reaction rate $dP/dtdq_1$ as a function of the momentum of the emitted particle $q_1$, both quantities are  
in atomic units. Here the value of dimensionless parameter $g''$ has
	been taken to be $10^{-5}$. 
        }
\label{fig:rate}
\end{figure}

We choose $k_m$ at the edge of a band with $\cos(k_m\xi)=1$. Since $k_i$ can be chosen arbitrarily we
can pick any value of $q_1$. For our calculation we set
 $q_1 = 600$. 
This value is chosen since it will lead to dominant contribution from 
$E_m$ values a little above the potential barrier $V_0$. This region
dominates the reaction rate. 
The value of $q_2$ is 
now given by
\begin{equation}
	q_2 = {E_i-E_f \over c} - q_1 
\end{equation}
This also fixes $k_f$ by the relationship $k_f = -q_1-q_2 + k_i-G-G'$. 
We note that
for each $E_m$ there are two wave functions, $\psi^+$ and $\psi^-$ which 
are complex conjugates of one another. Due to the delta function
$\delta_{ G,- k_m -  q_1 + k_i}$ only one of these would contribute for
a fixed $q_1$. 
The values of the real part of $I$ as a function of the
upper cutoff on the energy $E_m$ of the intermediate state is shown in  
Fig. \ref{fig:Ecutoffdep}. Here we show 
the contribution from dominant region $-a-2b<x<-a$. The contribution from
other regions is highly suppressed. We find that $I$ settles to a value 
of $-6.89\times 10^{-11}$ in atomic units.
The imaginary part behaves in a similar manner and settles to a value
approximately $2.6\times 10^{-11}$.
We also show the dependence of rate as a function of the emitted
$\phi$ particle momentum
$q_1$ in Fig. \ref{fig:rate} setting the coupling parameter $g''=10^{-5}$.
The total rate after integrating over $q_1$ is found to approximately
$2\times 10^{-13}$ in atomic units.

It is clear from Fig. \ref{fig:Ecutoffdep} that the second order process
is not suppressed. This is because the final value of $\bar I$
is comparable to the peak value obtained for some intermediate
value of cutoff Energy. The result would be negligible only if 
contributions from different energy eigenvalues show delicate cancellation, 
which does not
happen in this case. 
The behaviour of the amplitude is very similar to what is seen in our 
earlier paper for a different 3 dimensional model \cite{Ramkumar2022}. 
Here also a cancellation was seen which results in a small but significant
contribution. 

We may compare the rate with that obtained from 
the first order process. 
The rate in this case in given by
\begin{equation}
	{dP\over dt} = {g^2\over 2\hbar c\omega} \Bigg|\int dx \psi_f^* 
	e^{-iqx} \psi_i\Bigg|^2
\end{equation}
where $w$ and $q$ are the frequency and momentum of the emitted $\phi$ particles. 
The comparison depends on the 
choice of the dimensionless parameter $g''$.
For $g''=10^{-5}$, chosen earlier, we find that the rate is of order $10^{-76}$
in atomic units. Here we have chosen the same initial and final states as
for the second order process. The dominant suppression arises from the 
exceedingly small factor corresponding to the barrier penetration probability.
It is clear that the first 
order process is highly suppressed compared to the second order process, unless
we choose exceedingly small values for the dimensionless coupling parameter
$g''$.

\section{Conclusions}
In this paper we have studied the rate of nuclear fusion at low energies
using a $1+1$ dimensional lattice model. The model is a generalization
of the standard Kronig-Penney model and incorporates the deep nuclear 
potential. We consider fusion of a particle, which is analogous to 
proton, with a lattice site. The fusion rate of the standard first 
order process is very strongly suppressed due to tunnelling barrier.
We consider the rate at second order in perturbation theory. 
For this
case the rate can be large in some special circumstances, as described 
in earlier papers \cite{Jain2020,Jain2021,Ramkumar2022}. 
In \cite{Ramkumar2022} we had considered a three dimensional toy model
assuming a step potential. Here we found that the rate was very small if
the eigenfunctions show a continuous energy spectrum, which is achieved
if the potential vanishes at large distances. However it was argued
that in a medium the potential need not be zero at large distances 
and spectrum is not necessarily continuous. We modelled this phenomenon
by assuming an infinite potential barrier beyond a certain distance.
This lead to a significant rate for the fusion process \cite{Ramkumar2022}.
Here we 
are interested in determining whether a lattice model leads to an observable
rate. 

The fusion process takes place by emission of two particles, one at 
each interaction. We have tried a few different models for the interaction.
The simplest model in which the interaction Hamiltonian is taken to be
of the form given in Eq. \ref{eq:HI} with the operator ${\cal O}$ equal 
to unity we find that the rate is very small. The actual value is difficult
to obtain due to a very delicate cancellation among amplitudes due to different
intermediate states and we do not pursue this
in detail in the present paper. A modified interaction with the
operator ${\cal O}$ having a momemtum dependence given in Eq. \ref{eq:operator} is found to give a substantial rate. Although many interactions depend 
on momentum, it is not known whether such an interaction can arise 
in a real physical situation.
We have suggested that it might arise as an effective interaction or
a form factor if in the initial stage the process happens through the
Hydrogen atom rather than a proton. However, we have not examined this
possibility in detail.
Nevertheless, our result demonstrates
that quantum mechanics does not prohibit a relatively large fusion rate
even at very low energies and high potential barrier. 
To be precise, there does not seem to be any general rule that the
amplitude necessarily has to be small exceedingly small even for the second
order process. 

Our results show that nuclear fusion may take place in a crystalline 
lattice but requires special conditions. A detailed study is required
to determine whether
those may be met in a real physical situation. An alternative 
possibility is presented by
 a disordered system.  In such a system, all
states up to a certain energy level are localized. 
Hence, in a given
localized region, relevant for the fusion process, dominant contributions
would be obtained only from discretized energy states. 
 Hence we expect that fusion rate be a significant in
this case, in analogy with results obtained in \cite{Ramkumar2022}. 
A detailed calculation in a disordered physical system is postponed to
future research.

\bigskip
\noindent
 {\bf Acknowledgements:} We are very grateful to Amit Agarwal for useful
 discussions. 

\bibliographystyle{ieeetr}
\bibliography{nuclear}

\begin{thebibliography}{10}

\bibitem{doi:10.1002/9781118043493.ch41}
S.~B. Krivit, {\em Development of Low-Energy Nuclear Reaction Research},
  ch.~41, pp.~479--496.
\newblock John Wiley \& Sons, Ltd, 2011.

\bibitem{doi:10.1002/9781118043493.ch42}
L.~I. Urutskoev, {\em Low-Energy Nuclear Reactions: A Three-Stage Historical
  Perspective}, ch.~42, pp.~497--501.
\newblock John Wiley \& Sons, Ltd, 2011.

\bibitem{doi:10.1002/9781118043493.ch43}
M.~Srinivasan, G.~Miley, and E.~Storms, {\em Low-Energy Nuclear Reactions:
  Transmutations}, ch.~43, pp.~503--539.
\newblock John Wiley \& Sons, Ltd, 2011.

\bibitem{StormsCS2015}
E.~Storms, ``Introduction to the main experimental findings of the lenr
  field,'' {\em Current Science}, vol.~108, pp.~535--539, 02 2015.

\bibitem{mckubre2015cold}
M.~C. McKubre, ``Cold fusion: comments on the state of scientific proof,'' {\em
  Current Science}, pp.~495--498, 2015.

\bibitem{articleBi}
J.-P. Biberian, ``Biological transmutations,'' {\em Current Science}, vol.~108,
  pp.~633--635, 02 2015.

\bibitem{SRINIVASAN2020233}
M.~Srinivasan and K.~Rajeev, ``Chapter 13 - transmutations and isotopic shifts
  in lenr experiments,'' in {\em Cold Fusion} (J.-P. Biberian, ed.), pp.~233 --
  262, Elsevier, 2020.

\bibitem{Mizuno19}
T.~Mizuno and J.~Rothwell, ``Excess heat from palladium deposited on nickel,''
  {\em Journal of Condensed Matter Nuclear Science}, vol.~29, pp.~21--33, 2019.

\bibitem{krivit2009new}
S.~B. Krivit and J.~Marwan, ``A new look at low-energy nuclear reaction
  research,'' {\em Journal of Environmental Monitoring}, vol.~11, no.~10,
  pp.~1731--1746, 2009.

\bibitem{SinhaCS2015}
K.~P. Sinha, ``Model of low energy nuclear reactions in a solid matrix with
  defects,'' {\em Current Science}, vol.~108, pp.~516--518, 02 2015.

\bibitem{LiangCS2015}
C.~L. Liang, Z.~M. Dong, and X.~Z. Li, ``Selective resonant
  tunnelling–turning hydrogen-storage material into energetic material,''
  {\em Current Science}, vol.~108, pp.~519--523, 02 2015.

\bibitem{Hagelstein15}
P.~Hagelstein, ``Deuterium evolution reaction model and the fleischmann–pons
  experiment,'' {\em Journal of Condensed Matter Nuclear Science}, vol.~16,
  pp.~46--63, 02 2015.

\bibitem{Celani17}
F.~Celani, A.~Tommaso, and G.~Vassallo, ``The zitterbewegung interpretation of
  quantum mechanics as theoretical framework for ultra-dense deuterium and low
  energy nuclear reactions,'' {\em Journal of Condensed Matter Nuclear
  Science}, vol.~24, pp.~32--41, 02 2017.

\bibitem{kalman2019low}
P.~K{\'a}lm{\'a}n and T.~Keszthelyi, ``On low-energy nuclear reactions,'' {\em
  arXiv preprint arXiv:1907.05211}, 2019.

\bibitem{SPITALERI2016275}
C.~Spitaleri, C.~Bertulani, L.~Fortunato, and A.~Vitturi, ``The electron
  screening puzzle and nuclear clustering,'' {\em Physics Letters B}, vol.~755,
  pp.~275 -- 278, 2016.

\bibitem{Meulenberg19}
J.-L. Paillet and A.~Meulenberg, ``On highly relativistic deep electrons,''
  {\em Journal of Condensed Matter Nuclear Science}, vol.~29, pp.~472--492, 02
  2019.

\bibitem{PhysRevC.99.054620}
P.~K\'alm\'an and T.~Keszthelyi, ``Forbidden nuclear reactions,'' {\em Phys.
  Rev. C}, vol.~99, p.~054620, May 2019.

\bibitem{Jain2020}
P.~Jain, A.~Kumar, R.~Pala, and K.~P. Rajeev, ``Photon induced low-energy
  nuclear reactions,'' {\em Pramana}, vol.~96, no.~96, 2022.

\bibitem{Jain2021}
P.~Jain, A.~Kumar, K.~Ramkumar, R.~Pala, and K.~P. Rajeev, ``Low energy nuclear
  fusion with two photon emission,'' {\em JCMNS}, vol.~35, p.~1, 2021.

\bibitem{Ramkumar2022}
K.~Ramkumar, H.~Kumar, and P.~Jain, ``A toy model for low energy nuclear
  fusion,'' {\em Pramana}, vol.~97, no.~109, 2023.

\bibitem{merzbacher1998quantum}
E.~Merzbacher, {\em Quantum Mechanics}.
\newblock Wiley, 1998.

\bibitem{sakurai1967advanced}
J.~Sakurai, {\em Advanced Quantum Mechanics}.
\newblock Always learning, Pearson Education, Incorporated, 1967.

\bibitem{1968psen.book.....C}
D.~D. {Clayton}, {\em {Principles of stellar evolution and nucleosynthesis}}.
\newblock The University of Chicago Press, Chicago, 1968.

\bibitem{Ashcroft76}
N.~W. Ashcroft and N.~D. Mermin, {\em {S}olid {S}tate {P}hysics}.
\newblock Holt-Saunders, 1976.

\bibitem{doi:10.1142/7663}
E.~Abrahams, {\em 50 Years of Anderson Localization}.
\newblock WORLD SCIENTIFIC, 2010.

\bibitem{RevModPhys.57.287}
P.~A. Lee and T.~V. Ramakrishnan, ``Disordered electronic systems,'' {\em Rev.
  Mod. Phys.}, vol.~57, pp.~287--337, Apr 1985.

\bibitem{Biberian19}
J.-P. Biberian, ``Anomalous isotopic distribution of silver in a palladium
  cathode,'' {\em Journal of Condensed Matter Nuclear Science}, vol.~29,
  pp.~211--218, 2019.

\end{thebibliography}
\end{document}